\documentclass{article}

\usepackage{arXiv}

\usepackage[utf8]{inputenc} 
\usepackage[T1]{fontenc}    

\usepackage{xcolor}
\usepackage[colorlinks = true,
            linkcolor = gray,
            urlcolor  = gray,
            citecolor = gray,
            anchorcolor = red]{hyperref}
            
\usepackage{url}            

\usepackage{booktabs}       
\usepackage{multirow}       
\usepackage{tabularx}       
\usepackage{makecell}       
\newcolumntype{C}{>{\centering\arraybackslash}X} 

\usepackage{amsmath}        
\usepackage{amsfonts}       
\usepackage{nicefrac}       

\usepackage{microtype}      

\usepackage{apacite} 
\usepackage{doi}            

\usepackage{graphicx}       
\usepackage{float}          

\usepackage{fancyhdr}       
\title{
Investigating the use of terrain-following coordinates in AI-driven precipitation forecasts
    \thanks{\textit{Citation}: Sha, Y. et al. Investigating the contribution of terrain-following coordinates and conservation schemes in AI-driven precipitation forecasts. Submitted for publication in Geophysical Research Letters}
    \author{
      \textbf{Yingkai Sha}$^{\dagger}$,
      \textbf{John S. Schreck}$^{\dagger}$,
      \textbf{William Chapman}$^{\ddagger}$,
      \textbf{David John Gagne II}$^{\dagger}$\\[1em]
      Computational and Information Systems (CISL) Laboratory$^{\dagger}$ \\
      Climate and Global Dynamics (CGD) Laboratory$^{\ddagger}$ \\[1em]
      NSF National Center for Atmospheric Research \\
      Boulder, Colorado, USA\\[1em]
      \texttt{\{ksha, schreck, wchapman, dgagne\}@ucar.edu}
    }
}

\begin{document}
\maketitle

\begin{abstract}
Artificial Intelligence (AI) weather prediction (AIWP) models often produce ``blurry'' precipitation forecasts. This study presents a novel solution to tackle this problem---integrating terrain-following coordinates into AIWP models. Forecast experiments are conducted to evaluate the effectiveness of terrain-following coordinates using FuXi, an example AIWP model, adapted to 1.0$^\circ$ grid spacing data. Verification results show a largely improved estimation of extreme events and precipitation intensity spectra. Terrain-following coordinates are also found to collaborate well with global mass and energy conservation constraints, with a clear reduction of drizzle bias. Case studies reveal that terrain-following coordinates can represent near-surface winds better, which helps AIWP models in learning the relationships between precipitation and other prognostic variables. The result of this study suggests that terrain-following coordinates are worth considering for AIWP models in producing more accurate precipitation forecasts.
\end{abstract}

\section*{Plain Language Summary}
Artificial intelligence (AI) weather prediction (AIWP) models can forecast precipitation but often produce ``blurry'' results---overestimating drizzle while underestimating heavy rainfall. We aim to reduce this blurriness problem by developing an AIWP model using a hybrid sigma-pressure vertical coordinate system that follows the terrain, rather than pressure coordinates that intersect with terrain. We also developed and applied physical constraints under this coordinate system. We compare the results from this new AIWP model with its baseline versions and find that the new version can make more accurate precipitation forecasts with an improved distribution of precipitation values from light to heavy precipitation events. We suggest that it is beneficial for AIWP models to incorporate terrain-following coordinates in producing more accurate precipitation forecasts.

\section{Introduction}
Artificial Intelligence (AI) weather prediction (hereafter AIWP) models have achieved great success, driven by advances in neural network architectures, computational devices, and the availability of high-quality reanalysis datasets, such as the European Centre for Medium-Range Weather Forecasts (ECMWF) Reanalysis version 5 \cite<ERA5; >{hersbach2020era5}. State-of-the-art AIWP models can generate forecasts with low computational costs. Their forecast skill scores are competitive with top global Numerical Weather Prediction (NWP) models and can be improved further with additional training data \cite<e.g.>{bi2023accurate,lam2023learning,chen2023fuxi,nguyen2023climax,bonev2023spherical,bodnar2024aurora,nguyen2023scaling,willard2024analyzing,lang2024aifs,schreck2024community}. However, AIWP models exhibit lower fidelity in forecasting precipitation compared to prognostic, free-atmosphere variables. Recent evaluations on Weatherbench2 reveal that AIWP models tend to produce ``blurry'' precipitation forecasts. In many cases, they are verified to be worse than the ECMWF Integrated Forecast System (IFS) \cite{rasp2024weatherbench}.

To tackle the problem of AI-driven precipitation forecasts, this study proposes a novel solution---integrating terrain-following coordinates into AIWP models. Terrain-following coordinates have been used in NWP models since \citeA{phillips1957coordinate}. Among these, hybrid sigma-pressure coordinates-- which follow terrain near the surface and transition to constant pressure levels aloft --have seen the widest adoption \cite{simmons1981energy,ritchie1991application,ecmwf2016dynamics}. As discussed in \citeA{rasp2024weatherbench}, an important reason for the blurry AI-driven precipitation forecasts is the lack of representation of local-scale features. Here, we hypothesize that hybrid sigma-pressure coordinates have the potential to represent near-surface weather conditions better and help AIWP models understand the relationships between local-scale information and precipitation characteristics when making forecasts.

Several studies have identified the effectiveness of using physical constraints in AIWP models \cite<e.g.>{wang2025condensnet,watt2023ace,yuval2024neural,sha2025improving}. Many of which also benefit precipitation forecasts. Notably, in \citeA{sha2025improving}, the integration of global mass and energy conservation constraints in constant pressure level AIWP models provided a clear reduction of drizzle bias---a common problem of weather models \cite<e.g.>{gutowski2003temporal}. To ensure that the benefits of physical constraints can be preserved under terrain-following coordinates, in this study, we developed a new set of conservation schemes as constraints, similar to that of \citeA{sha2025improving}, but specifically for hybrid sigma-pressure levels to examine its interaction with the terrain-following coordinates.

We evaluate the proposed solutions above using FuXi \cite{chen2023fuxi}, an AIWP model architecture validated on Weatherbench2 \cite{rasp2024weatherbench}. In our experiments, the ERA5, re-gridded to 1.0$\mathrm{^\circ}$ grid spacing, is used for training, while the Integrated Multi-satellitE Retrievals for Global Precipitation Measurement \cite<IMERG; >{huffman2020integrated} serves as the verification target. The technical foundation of this study is based on the Community Research Earth Digital Intelligence Twin \cite<CREDIT; >{schreck2024community} platform hosted at NSF National Center for Atmospheric Research (NCAR).

This study addresses three key questions: (1) Can terrain-following coordinates improve AI-driven precipitation forecasts?  (2) Can AIWP physical constraints be applied under terrain-following coordinates and provide combined precipitation forecast benefits? (3) Can we provide a physical basis for these improvements? By answering the questions, we aim to bring insights into the limitations of constant pressure level AIWP models in precipitation forecasts and how such limitations can be tackled. In addition, we hope to inspire future innovations to introduce more atmospheric science domain knowledge to the development of AIWP models.

\section{Data and methods}\label{sec2}

\subsection{Data}\label{sec21}

\begin{table}
\begin{center}
\caption{Input and output variables of the AIWP models.}\label{tab1}
\renewcommand{\arraystretch}{1.2}
\begin{tabularx}{\textwidth}
{c >{\centering\arraybackslash}X c c}
\specialrule{1.5pt}{0pt}{3pt}
Type & Variable Name & Units & Role \\ 
\midrule
\multirow{5}{*}{Upper air\textsuperscript{a}}
& Zonal Wind                                 & $\mathrm{m \cdot s^{-1}}$   & \multirow{5}{*}{Prognostic, Instantaneous} \\
& Meridional Wind                            & $\mathrm{m \cdot s^{-1}}$   &\\
& Air Temperature                            & $\mathrm{K}$                &\\
& Specific Total Water\textsuperscript{b}    & $\mathrm{kg \cdot kg^{-1}}$ &\\
& Geopotential\textsuperscript{c}     & $\mathrm{m^2\cdot s^{-2}}$                &\\
\midrule
\multirow{4}{*}{Single level}& Mean Sea Level Pressure\textsuperscript{d} & $\mathrm{Pa}$ & \multirow{4}{*}{Prognostic, Instantaneous}\\
& Surface Pressure\textsuperscript{e}        & $\mathrm{Pa}$                & \\
& 2-Meter Temperature                        & $\mathrm{K}$                & \\
& 10-Meter Zonal Wind                        & $\mathrm{m \cdot s^{-1}}$   &\\
& 10-Meter Meridional Wind                   & $\mathrm{m \cdot s^{-1}}$   &\\
\midrule
\multirow{9}{*}{Flux form\textsuperscript{f}} & Total Precipitation          & $\mathrm{m}$                & \multirow{8}{*}{Diagnostic, Cumulative} \\
& Evaporation                                & $\mathrm{m}$                & \\
& Top-of-atmosphere Net Solar Radiation      & $\mathrm{J \cdot m^{-2}}$   &\\
& Outgoing Longwave Radiation                & $\mathrm{J \cdot m^{-2}}$   &\\
& Surface Net Solar Radiation                & $\mathrm{J \cdot m^{-2}}$   &\\
& Surface Net Longwave Radiation             & $\mathrm{J \cdot m^{-2}}$   &\\
& Surface Net Sensible Heat Flux           & $\mathrm{J \cdot m^{-2}}$   &\\
& Surface Net Latent Heat Flux             & $\mathrm{J \cdot m^{-2}}$   &\\
\cmidrule(lr){2-4}
& Top-of-atmosphere Incident Solar Radiation & $\mathrm{J \cdot m^{-2}}$   & Input-only, Cumulative \\
\midrule
\multirow{4}{*}{Others}                  & Sea-ice Cover                              & n/a                         & Input-only, Instantaneous \\
& Geopotential at the Surface                & $\mathrm{m^2 \cdot s^{-2}}$ & Input-only, Static \\
& Land-sea Mask                              & n/a                         & Input-only, Static \\
& Soil Type                                  & n/a                         & Input-only, Static \\
\specialrule{1.5pt}{3pt}{0pt}
\end{tabularx}
\end{center}
\textsuperscript{a} Upper air variables are prepared separately on constant pressure levels and hybrid sigma-pressure levels.\\
\textsuperscript{b} Specific total water is the combination of specific humidity, cloud liquid water content, and rainwater content.\\
\textsuperscript{c} Geopotential is available for constant pressure level models only.\\
\textsuperscript{d} Mean sea level pressure is available for constant pressure level models only.\\
\textsuperscript{e} Surface pressure is available for hybrid sigma-pressure level models only.\\
\textsuperscript{f} Flux form variables are accumulated every 6 hours. Downward flux is positive.
\end{table}

In this study, the AIWP models are trained using the ERA5. Two sets of upper-air variables are prepared: one on hybrid sigma-pressure levels and one on constant pressure levels (see Table~\ref{tab1}). The hybrid sigma-pressure dataset uses surface pressure as its coordinate reference and excludes geopotential height, while the constant pressure dataset incorporates mean sea level pressure. The two coordinates share the same flux form variables and input-only variables. Their specific humidity and the liquid phase moisture are combined as specific total water. Each ERA5 version is pre-processed to 1.0$^\circ$ grid spacing with 6-hour intervals spanning 1 January 1979 to 31 December 2021.

For vertical dimensions, the constant pressure level dataset has 13 vertical levels on \{1, 50, 150, 200, 250, 300, 400, 500, 600, 700, 850, 925, 1000\} hPa. The hybrid sigma-pressure level dataset also has 13 vertical levels, selected from the \{15, 48, 68, 74, 79, 83, 90, 96, 101, 105, 114, 120, 133\}-th IFS full levels (Figure~\ref{fig1}a and b). From the mean sea level pressure of 1013.25 hPa, these hybrid sigma-pressure levels resemble their constant pressure level counterparts closely. Technical details of the ERA5 pre-processing and coordinate coefficients are provided in the Supporting Information.

IMERG is a satellite-based precipitation product developed by the National Aeronautics and Space Administration (NASA). In this study, IMERG is applied as verification targets only and is not used in the training of AIWP models (i.e., AIWP models are trained using the ERA5 total precipitation and have no access to IMERG). The IMERG version 7.0 final precipitation (L3) daily product \cite{huffman2020integrated} is converted to 1.0$^\circ$ grid spacing. The daily climatology of IMERG is computed from 2000 to 2019.

\subsection{FuXi with physical constraints}\label{sec22}

\begin{figure}
    \centering
    \includegraphics[width=\columnwidth]{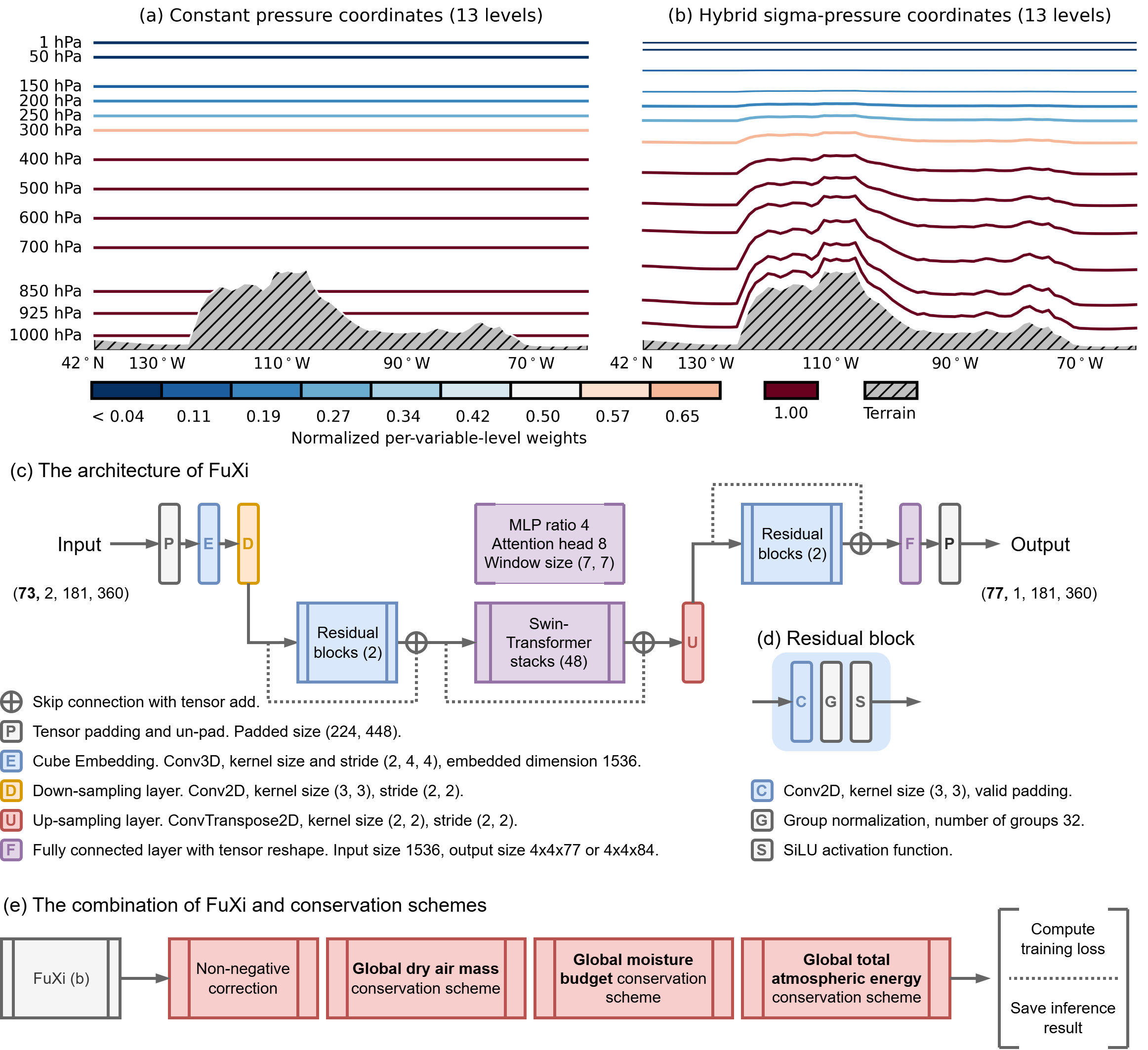}
    \caption{(a) The arrangement of constant pressure levels over a zonal cross-section of 135$^\circ$-65$^\circ$W on 42$^\circ$N. (b) As in (a) but for hybrid sigma-pressure levels. Line colors are the normalized per-variable-level weights. Higher weights mean larger influences in AIWP model training. (c) The architecture of FuXi with the I/O sizes of this study. (d) The architecture of the FuXi residual block.
    (e) The application of conservation schemes during the training and inference of FuXi.}
    \label{fig1}
\end{figure}

FuXi \cite{chen2023fuxi} is an operational AIWP model proposed for constant pressure level forecasts. It was also modified in \citeA{sha2025improving} and \citeA{schreck2024community} for research purposes. The base architecture of FuXi is Swin-Transformer, which has been adopted by a wide range of AIWP models \cite<e.g.>{willard2024analyzing,nguyen2023scaling}, making FuXi a good example to experiment with solutions of blurring precipitation forecasts.

The implementation of FuXi follows its original design [Figure~\ref{fig2}c and d; \citeA{chen2023fuxi}] with a few modifications. Tensor padding is applied to ensure that the 1.0$^\circ$ data is compatible with the FuXi architecture. Spectral normalization \cite{miyato2018spectral} is applied to all trainable layers to stabilize the model training. Model cascading is not used to avoid the inconsistencies of the forecast trajectories.

Conservation schemes are applied after the FuXi output layer (Figure~\ref{fig1}e). They provide (1) Nonnegative correction, (2) Global dry air mass conservation, (3) Global moisture budget conservation, and (4) Global total atmospheric energy conservation. The schemes are applied in the specified order in both model training and inference. Similar to \citeA{sha2025improving}, they are proposed to improve the physical consistency of AIWP models and guide them to produce forecasts that obey the conservation laws. Additional technical details of the FuXi model with conservation schemes are provided in the Supporting Information.

\subsection{Experiment design and model training}\label{sec23}

Four AIWP model runs are conducted and compared. Their names, configurations, and purposes are summarized below:

\begin{enumerate}
    \item \textit{FuXi-plevel-base} is the constant pressure level FuXi without any physical constraints. This configuration provides the most basic AIWP setup and serves as a performance reference in this experiment.

    \item \textit{FuXi-plevel-physics} is the constant pressure level FuXi with physical constraints. This configuration examines the contribution of physical constraints on precipitation.
    
    \item \textit{FuXi-sigma-base} is the hybrid sigma-pressure level FuXi without physical constraints. This configuration measures and isolates the contribution of terrain-following coordinates in precipitation forecasts.

    \item \textit{FuXi-sigma-physics} is the hybrid sigma-pressure level FuXi with physical constraints. By comparing FuXi-sigma-physics against FuXi-plevel-physics, the contribution of terrain-following coordinates can be quantified. Similarly, by comparing FuXi-sigma-physics against FuXi-sigma-base, the benefit of conservation schemes can be identified.
\end{enumerate}

All four FuXi configurations are trained using 32 NVIDIA A100 GPUs using Pytorch \cite{paszke2019pytorch}. The ERA5 dataset is divided into three parts, with 1979-2018 for model training, 2019 for validation, and 2020-2021 for verification.

The ECMWF IFS high-resolution forecast (IFS-HRES) is included as an NWP baseline. IFS-HRES is recognized as the best operational medium-range weather forecasting system. In Weatherbench2, it also showed competitive precipitation forecast performance, typically outperforming other blurry AIWP models \cite{rasp2024weatherbench}. The IFS-HRES precipitation forecasts are collected from the Weatherbench 2 data archive and interpolated to the 1.0$^\circ$ grid spacing for comparisons. Hereafter, this pre-processed version of IFS-HRES is referred to as ``IFS-HRES'' directly. 

\subsection{Verification methods}

All experiment members produce 6-hourly precipitation forecasts for up to 10 days. The 6-hourly forecasts are accumulated to daily values for verifications. The deterministic verification of precipitation is based on two categorical scores: Threat Score (TS) and Stable Equitable Error in Probability Space (SEEPS). We prefer not to include continuous metrics such as Root Mean Squared Error (RMSE) and mean bias because they favor ``blurry'' precipitation forecasts and may give AIWP models unfair advantages \cite<also explained in>{rasp2024weatherbench}.

TS is a classification metric that evaluates how well a forecast predicts the occurrence of an event \cite[page 263]{wilks2006statistical}. It is defined using the confusion matrix components of True Positive (TP), False Positive (FP), and False Negative (FN) samples:

\begin{equation}\label{equ_TS}
\mathrm{TP} = \frac{\mathrm{TP}}{\mathrm{TP}+\mathrm{FP}+\mathrm{FN}}
\end{equation}

\noindent
TS does not consider True Negative samples, which makes it a good metric for verifying precipitation events. In this study, TS is computed using the peak-over-threshold of 0.1 mm and 25 $\mathrm{mm\cdot day^{-1}}$ for drizzle and extreme events, respectively. For each threshold, two sets of TS are computed for (1) the entire globe and (2) land grid cells only.

The implementation of SEEPS follows \citeA{rodwell2010new} and \citeA{rasp2024weatherbench}. Precipitation forecasts and IMERG targets are converted to categories of ``dry'', ``light'', and ``heavy'' events using climatology-based percentiles. The converted categorical forecast is then verified using a 3-by-3 contingency table for each location, forecast lead time, and all initialization times. SEEPS is the matrix multiplication between the contingency table and the following score matrix:

\begin{equation}
S = \begin{pmatrix}
0 & \frac{1}{1-p} & \frac{4}{1-p} \\
\frac{1}{p} & 0 & \frac{3}{1-p} \\
\frac{1}{p}+\frac{3}{2+p} & \frac{3}{2+p} & 0
\end{pmatrix}
\end{equation}

\noindent
Where $p$ is the climatological probability of dry days. Grid cells with $p\in\left(0.1, 0.85\right)$ are included. The global weighted SEEPS is the cosine-latitude-weighted sum of SEEPS on all available grid cells.

Quantile values of precipitation forecasts and IMERG targets are compared to examine the distribution of precipitation intensities. This verification is similar to the use of quantile-quantile (q-q) plots \cite[page 152]{wilks2006statistical} but with the IMERG quantiles serving as the distribution reference. The zonal spectral energy of precipitation is also compared to examine the smoothness of forecasts. Its technical procedure generally follows \citeA{rasp2024weatherbench}, with the zonal average covering the entire globe. The two verifications above do not penalize positional bias, which means the overly smoothed forecasts of AIWP models will not be favored in this comparison.


\section{Results}\label{sec3}

\subsection{Verification scores}\label{sec31}

\begin{figure}
    \centering
    \includegraphics[width=\columnwidth]{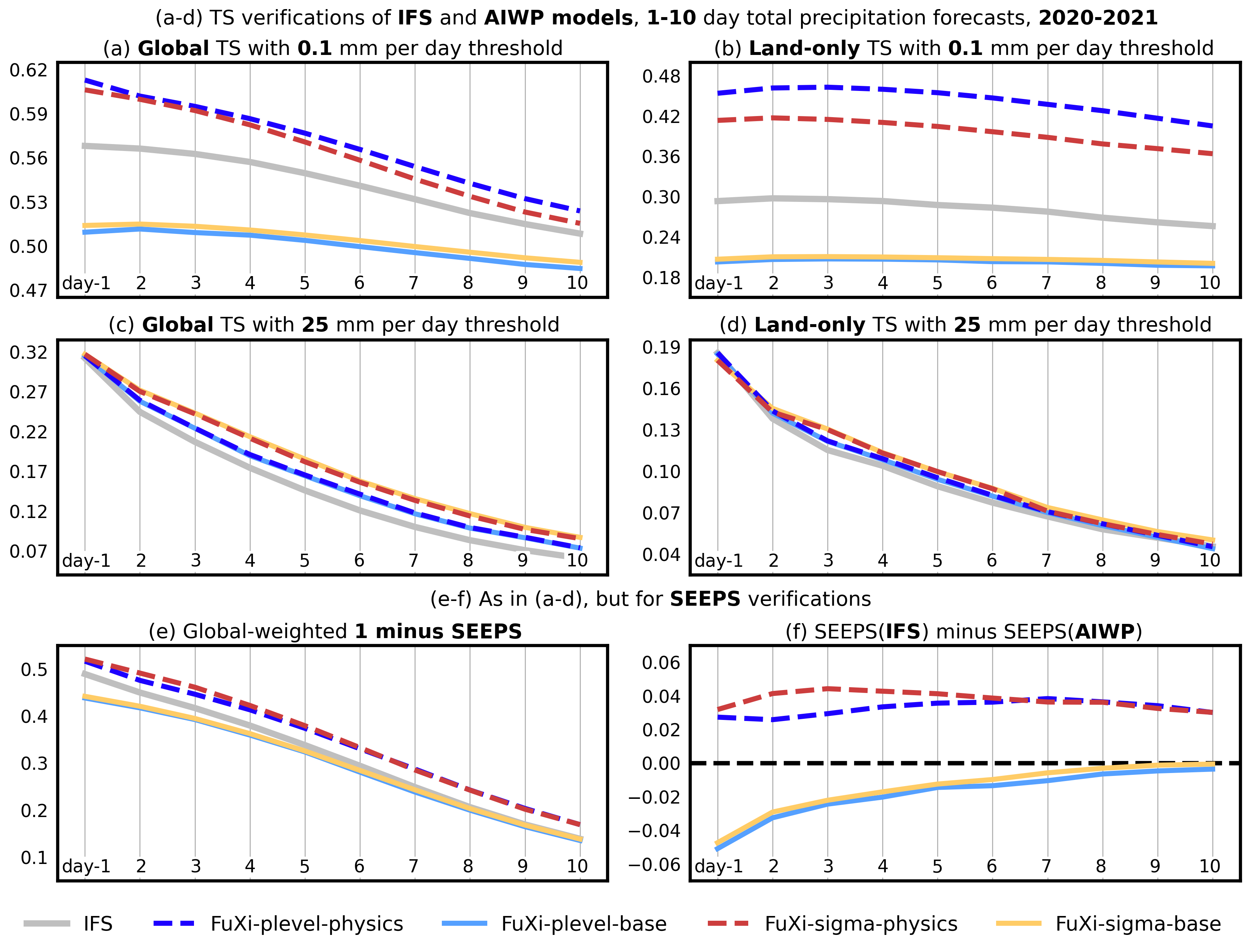}
    \caption{
    TS and SEEPS verification of daily total precipitation in IFS-HRES (gray solid line), FuXi-plevel-physics (blue dashed line), FuXi-plevel-base (cyan solid line), FuXi-sigma-physics (red dashed line), and FuXi-sigma-base (orange solid line) in 2020-2021. IMERG is the verification target. (a) Domain-wise TS by forecast lead times with 0.1 $\mathrm{mm\cdot day^{-1}}$ threshold. (b) as in (a), but for land grid cells only. (c) as in (a), but for 25 $\mathrm{mm\cdot day^{-1}}$ threshold. (d) as in (c), but for land grid cells only. (e) Global-weighted averaged SEEPS by forecast lead times. (f) The SEEPS of IFS-HRES in (e) minus the SEEPS of AIWP models. For (a-e), higher means better. In (f), higher means AIWP models are better than IFS-HRES.}
    \label{fig2}
\end{figure}

TS is applied to the deterministic verifications of precipitation forecasts using IMERG as targets. For the 0.1 $\mathrm{mm\cdot day^{-1}}$ TS, FuXi-plevel-physics and FuXi-sigma-physics perform well (Figure~\ref{fig2}a and b). IFS-HRES falls behind, while FuXi-plevel-base and FuXi-sigma-base perform the worst. This verification assesses the ability of AIWP models to predict 0.1 mm-level drizzle in the correct locations. The two base configurations likely overestimated the amount of drizzle due to their blurriness, which led to suboptimal TS performance. On the contrary, FuXi-plevel-physics and FuXi-sigma-physics perform better, indicating that using conservation schemes can reduce drizzle bias and improve the categorical verification of light precipitation forecasts. In addition, FuXi-plevel-physics outperformed FuXi-sigma-physics slightly, which is likely explained by the variation of their conservation schemes: FuXi-sigma-physics constraints surface pressure for the conservation of global dry mass (see Supporting Information for technical details), whereas FuXi-plevel-physics constraints the pressure level integral of humidity across all vertical layers \cite{sha2025improving}. The latter provides a stronger regularization effect on reducing the drizzle bias. Overall, the difference between FuXi-plevel-physics and FuXi-sigma-physics is minor, and the verification result confirms that the hybrid sigma-pressure level AIWP models can collaborate well with physical constraints and receive similar benefits compared to their constant pressure level counterparts.

The 25 $\mathrm{mm\cdot day^{-1}}$ TS is verified to examine the ability of AIWP models to predict heavy-to-extreme events (Figure~\ref{fig2}c and d). In this verification, IFS-HRES falls behind at short forecast lead times and catches up after day 7, especially in land areas. FuXi-sigma-physics and FuXi-sigma-base are comparably good, while FuXi-plevel-base and FuXi-plevel-physics are the worest. The contrast between FuXi-plevel-physics and FuXi-sigma-physics suggests that using terrain-following coordinates can improve extreme precipitation forecasts. In addition, the minimal difference between FuXi-sigma-physics and FuXi-sigma-base shows that the use of conservation constraints, although improved drizzle forecasts, is neutral to the performance change of extreme precipitation forecasts---it does not improve on extreme events, but also does not cause performance downgrades.

SEEPS is applied to measure the performance of precipitation forecasts that combine the skills of light and heavy precipitation events. Based on Figure~\ref{fig2}e, FuXi-sigma-physics, the combination of conservation schemes and terrain-following coordinates, provides promising results (Figure~\ref{fig2}f). FuXi-plevel-physics, although slightly behind FuXi-sigma-physics, is ranked as the second. This is likely because the good performance of FuXi-plevel-physics on light precipitation categories has outweighed its limitation in forecasting statistically rare heavy-to-extreme events. FuXi-plevel-base and FuXi-sigma-base have the worst SEEPS due to their poor performance on drizzle events.

\subsection{Distribution-based verification}\label{sec32}

\begin{figure}
    \centering
    \includegraphics[width=\columnwidth]{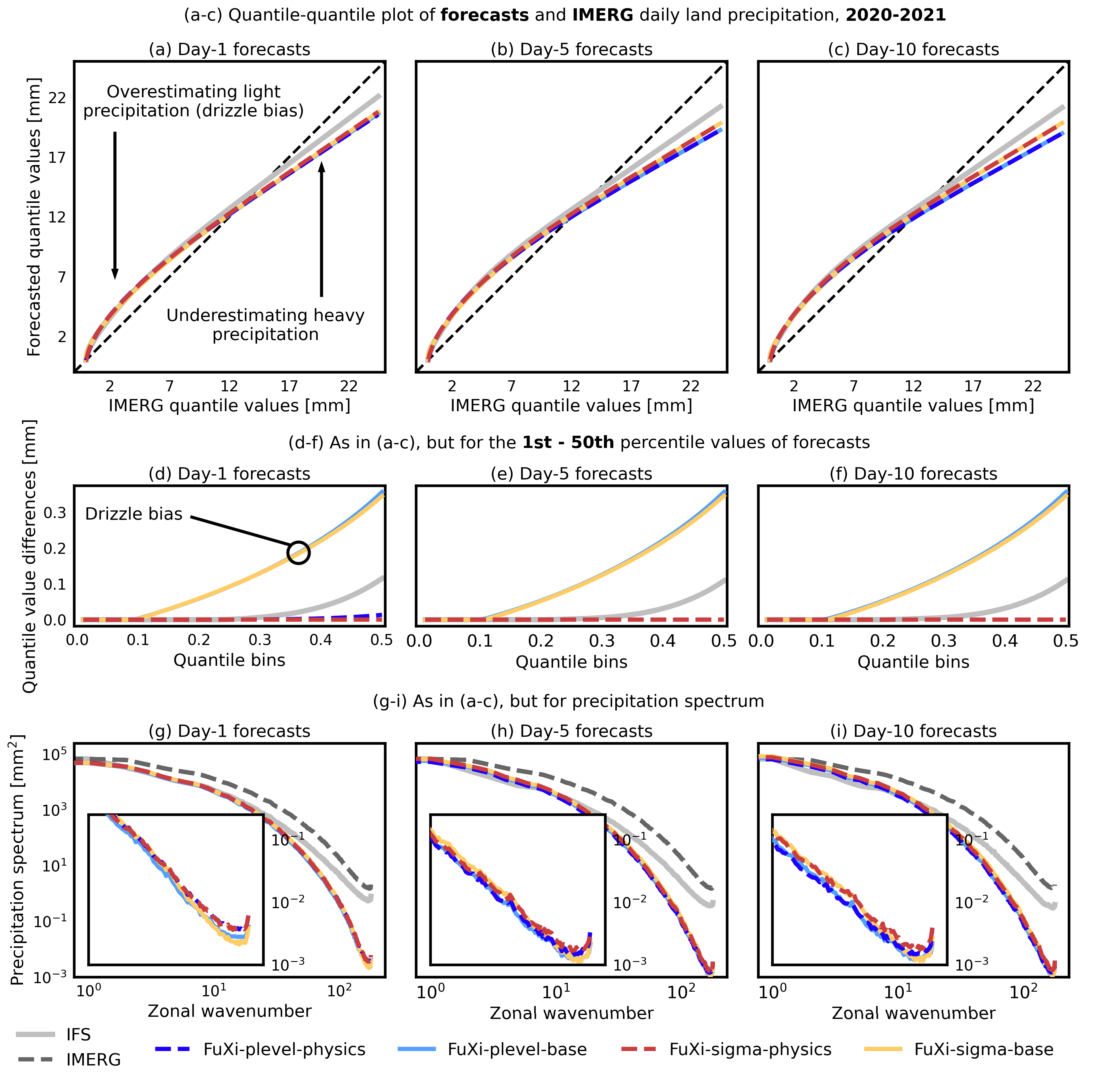}
    \caption{(a-c) Quantile-quantile comparisons between IMERG (x-axis) and forecasts (y-axis) in 2020-2021 and for 0.01 to 0.99 quantile bins. Participated members are IFS-HRES (gray solid line), FuXi-plevel-physics (blue dashed line), FuXi-plevel-base (cyan solid line), FuXi-sigma-physics (red dashed line), and FuXi-sigma-base (orange solid line). (a-c) are computed from day-1, day-5, and day-10 forecasts, respectively. (d-f) Comparisons of quantile values between forecasts and IMERG for 0.01 to 0.5 quantile bins. (g-i) The zonal energy spectra of precipitation forecasts and IMERG (gray dashed line). The zoom-ins cover the bottom right part of each panel. For (a-c), the diagonal line means a perfect agreement between IMERG and forecasted quantile values. In (d-f), positive means an overestimation of precipitation on a given quantile bin.}
    \label{fig3}
\end{figure}

In Figure~\ref{fig3}, quantile-based verification is presented to examine the distribution of precipitation intensities. Based on the q-q plots in Figure~\ref{fig3}a-c, all models underestimate the precipitation extremes. AIWP models exhibit stronger underestimations, and their forecasted extremes decrease with increasing forecast lead times (c.f. Figure~\ref{fig3}a and c). This result is similar to the findings in \citeA{radford2025comparison} in the United States. FuXi-plevel-base and FuXi-plevel-physics have the worst underestimation of extremes, whereas the IFS-HRES performed the best. FuXi-sigma-physics and FuXi-sigma-base fall behind IFS-HRES but are verified to be better than FuXi-plevel-physics, especially at longer forecast lead times. This further confirms that using terrain-following coordinates can improve extreme precipitation forecasts by fixing the right-side tail of precipitation distributions.

The difference of left-side quantiles between IMERG and precipitation forecasts is shown in Figure~\ref{fig3}d-f. The goal of this verification is to examine the drizzle bias in 0.01-0.5 quantile ranges from a Cumulative Distribution Function (CDF) perspective. This verification specifically highlights ``at which quantile the model starts forecasting drizzle''. Here, Fuxi-plevel-base and FuXi-sigma-base are verified to be the worst. Its overestimation of drizzle starts from the 10th percentile and increases rapidly on higher quantile bins. FuXi-plevel-physics and FuXi-sigma-physics perform the best by maintaining a minimal difference with IMERG. This verification confirms that the poor performance of FuXi-plevel-base and FuXi-sigma-base in the 0.1 mm $\mathrm{mm\cdot day^{-1}}$ TS is not due to positional bias of individual cases, but rather, a systematic overestimation of drizzle.

The zonal spectral energy of precipitation forecasts is verified in Figure~\ref{fig3}g-i. All forecasts underestimate the energy spectrum in high-frequency zones, pointing to blurrier forecasts compared to IMERG. This problem is especially profound for AIWP models, with all FuXi runs producing highly similar spectral energy. Under zoomed-in panels, however, some distinguishable energy distribution characteristics can still be identified: FuXi-sigma-base and FuXi-sigma-physics produced forecasts with higher spectral energy, indicating a mitigation of the bluriness problem. On the ``J''-shaped tail of the precipitation spectrum, Fuxi-plevel-physics and Fuxi-sigma-physics produced higher spectral energy. This is explained by the better separation of dry and light precipitation areas across drizzle events.

\subsection{A case study}\label{sec33}

\begin{figure}
    \centering
    \includegraphics[width=\columnwidth]{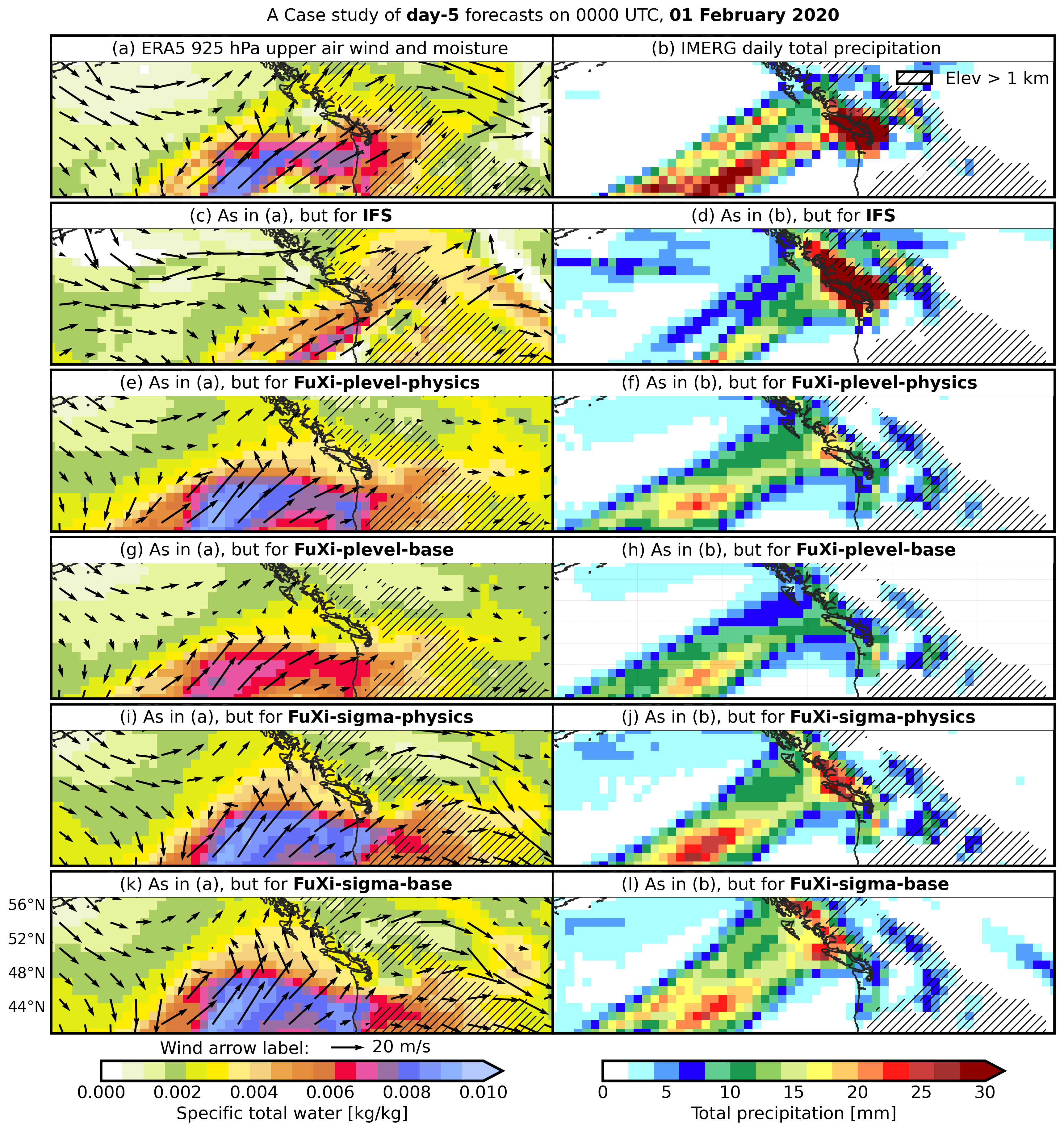}
    \caption{An example case on 0000 UTC, 1 February 2020, the West Coast of North America. (a) 925 hPa specific total water (color shades) and horizontal winds (arrows) in the ERA5. (b) Daily precipitation in IMERG. (c, d) As in (a, b), but for the day-5 forecast of IFS-HRES. (e, f) As in (c, d), but for FuXi-plevel-physics. (g, h) As in (c, d), but for FuXi-plevel-base. (i, j) As in (c, d), but for FuXi-sigma-physics on its 12th hybrid sigma-pressure level. (k, l) As in (I, J), but for FuXi-sigma-base. Hatched regions have terrain elevation higher than 1000 m.}
    \label{fig4}
\end{figure}

An atmospheric river event on 0000 UTC, 1 February 2020, is presented to explain why hybrid sigma-pressure level AIWP models can forecast precipitation extremes better. Based on the 925 hPa level atmospheric condition in ERA5, a large, moist air mass formed and approached the Pacific Northwest with strong westerly winds (Figure~\ref{fig4}a). The IMERG shows that extreme precipitation has occurred in the South Vancouver Island-Lower Mainland area (Figure~\ref{fig4}b).

The daily precipitation forecasts of AIWP models generally captured the correct locations 5 days prior. However, FuXi-plevel-base and FuXi-plevel-physics clearly underestimate the precipitation intensity of this event. Compared to IMERG, this underestimation is roughly 15 mm on the windward side of coastal mountains (Figure~\ref{fig4}f and h). FuXi-sigma-physics and FuXi-sigma-base produced stronger precipitation rates, which align more closely with IMERG. IFS-HRES exhibits large errors over the Pacific Ocean, but it also estimates the intensity of precipitation centers well, which means FuXi-plevel-base and FuXi-plevel-physics, the two constant pressure level AIWP models, performed the worst on this heavy-to-extreme event.

The forecasted near-surface states can explain the performance difference. In FuXi-plevel-base and FuXi-plevel-physics, 925 hPa constant pressure level horizontal wind speeds are largely reduced over the coastal mountains (c.f. Figure~\ref{fig4}a and e over the hatched area). In comparison, wind fields on the 12th level (i.e., the hybrid sigma-pressure level that is roughly on the 925 hPa pressure plane) of FuXi-sigma-physics and FuXi-sigma-base have a smoother transition from ocean to coastal mountains. We speculate that the different representations of near-surface winds in the two coordinates may have impacted the performance difference in AIWP precipitation forecasts. All AIWP models produced highly similar moisture distributions. However, in constant pressure coordinates, the 925 hPa westerly winds intersect with orography, and the ``below ground'' interpolation of constant pressure level vertical coordinate results in a less accurate representation of the near-surface circulation. This can mislead AIWP models in downplaying the impact of lower-atmosphere moisture transport. That said, a key factor that makes hybrid sigma-pressure level AIWP models excel in extreme precipitation forecasts is their ability to represent near-surface conditions more accurately, which benefits AIWP models in capturing the relationships between precipitation extremes and other lower-atmosphere prognostic variables.

\section{Discussion and conclusions}\label{sec4}

The visually blurry AI-driven precipitation forecasts raise two primary issues: drizzle bias and the underestimation of precipitation extremes. This study proposed solutions that can tackle both issues: (1) using terrain-following coordinates to improve extreme precipitation forecasts, and (2) applying global mass and energy conservation constraints under terrain-following coordinates to reduce drizzle bias. 

Numerical experiments confirmed the effectiveness of using terrain-following coordinates. FuXi-sigma-physics, the AIWP configuration that incorporates the full solution, provides the best precipitation forecasts with reduced drizzle bias and largely improved Threat Score (TS) on 25 $\mathrm{mm\cdot day^{-1}}$ threshold--- outperforming the constant pressure level AIWP models on extreme precipitation forecasts. 

A case study of an atmospheric river event is conducted to investigate the contribution of terrain-following coordinates. It reveals that terrain-following coordinates can represent near-surface wind fields better over land areas, whereas on constant pressure planes, due to the ``below ground'' interpolation, near-surface winds may be represented poorly, causing AIWP models to overlook their importance in precipitation forecasts. This shows that by providing better near-surface dynamics, terrain-following coordinates can help AIWP models learn precipitation-related information better and improve the extreme forecasts.

The solution of this study, using terrain-following coordinates, is inspired by the atmospheric science domain knowledge, and with a specific aim of improving precipitation forecasts. FuXi is the example AIWP architecture, but the solution is not tied to a specific neural network design. State-of-the-art AIWP models can maintain their constant pressure level structure while allocating more channels to represent near-surface dynamics under terrain-following coordinates. New AIWP models can also be developed on terrain-following coordinates directly. In addition, this study leverages 1.0$^\circ$ data to develop AIWP models, which allows four AIWP models to be trained effectively under limited computational resources. Based on the systematic comparisons of AIWP models, this study would leave a key remark on the future development of higher-resolution, multi-purpose AIWP models.

In conclusion, utilizing terrain-following coordinates and incorporating them with global mass and energy conservation constraints can improve AI-driven precipitation forecasts for both drizzle and heavy-to-extreme events. A wide range of AIWP models can benefit from this change, producing more skillful precipitation forecasts across various application scenarios.

%
%

\section*{Supporting Information}

\setcounter{figure}{0}
\setcounter{table}{0}
\renewcommand{\thefigure}{\arabic{figure}S}
\renewcommand{\thetable}{\arabic{table}S}

\noindent\textbf{Introduction}

This Supporting Information provides extended details of the ERA5 data pre-processing (main article Section 2.1), the FuXi architecture (main article Section 2.2), global mass and energy conservation constraints in hybrid sigma-pressure AIWP models (main article Section 2.2), and AIWP model training (main article Section 2.3). Additional verification results are also provided, covering the verification of AIWP prognostic variable forecasts and two extended example case studies. 

\noindent\textbf{S1. Extended details of ERA5 data access and pre-processing}

ERA5 is a global reanalysis dataset produced using the European Centre for Medium-Range Weather Forecasts (ECMWF) Integrated Forecasting System (IFS) version Cy41r2 \cite{hersbach2020era5}. At ECMWF, the ERA5 is first produced on 137 hybrid sigma-pressure levels and then post-processed to 37 constant pressure levels. In this study, both ERA5 versions are used. A full list of variables is provided in the Table~1 of the main article.

We collect the ERA5 from two sources: the NSF NCAR, Research Data Archive \cite{ecmwf2019era5}, and the Google Research, Analysis-Ready, Cloud Optimized (ARCO) ERA5 \cite{carver2023arcoera5}. The original data is hourly and has 0.25$^\circ$ grid spacing. We re-gridded them to 1.0$^\circ$ using conservative interpolation, similar to that of Weatherbench2 \cite{rasp2024weatherbench}. Hourly quantities are converted to 6-hourly following the validity time convention of ECMWF. Instantaneous variables are selected based on the ending time of every 6 hours; flux form variables are accumulated within each 6-hour time window. For vertical dimensions, constant pressure variables are selected on \{1, 50, 150, 200, 250, 300, 400, 500, 600, 700, 850, 925, 1000\} hPa levels; hybrid sigma-pressure level variables are selected from the \{15, 48, 68, 74, 79, 83, 90, 96, 101, 105, 114, 120, 133\}-th IFS full levels; their corresponding half-level coefficients are provided in Table~\ref{tab2}. The selected levels are also visualized in the Figure~1 of the main article. 

All variables in the pre-processed 1.0$^\circ$, 6-hourly datasets, except sea-ice cover, land-sea mask, and soil type, are normalized using z-score. Their mean and standard deviation are computed in 1979-2018. Soil type is rescaled from categorical integer values to 0.0-1.0 float numbers. The land-sea mask and the sea-ice cover are combined, with pure land, ocean, and sea-ice having float numbers of 1.0, 0.0, and -1.0, respectively. The steps above are similar to that of \citeA{sha2025improving}, but with mean and standard deviation values derived separatly from hybrid sigma-pressure and constant pressure variables.

\noindent\textbf{S2. Extended technical details of FuXi}

In this study, the architecture and hyperparameters of FuXi are aligned with its original design in \citeA{chen2023fuxi}. The cube embedding layer has patch sizes of (2, 4, 4) for the time, latitude, and longitude dimensions, respectively. The embedded hidden dimension is 1536. The U-Transformer of FuXi consists of tensor resampling and residual blocks using 2-D convolutional layers. The core of the U-Transformer is a 48 stack of Swin-Transformer V2 blocks. Each block is configured with 8 attention heads, window sizes of (7, 7), and MLP ratios of 4. The above technical details are also summarized in Figure~1c and d of the main article.

Similar to \citeA{sha2025improving} and \citeA{schreck2024community}, several changes are applied to the original FuXi design. Model cascading is not applied. This choice can cause inconsistencies when switching the models, and the problem is particularly profound for precipitation because the cascaded FuXi is optimized using mean squared error, whereas precipitation forecasts are verified using categorical scores.

Boundary padding is applied to ensure that the FuXi design is compatible with the 1.0$^\circ$ data. FuXi has 4-by-4 patch embedding followed by 2-by-2 tensor re-sampling and 7-by-7 shift window sizes. Thus, its compatible tensor size must be $N\times 56$. This study pads the input tensor size from (181, 360) to (224, 448) using a geometry-informed approach. It performs circular padding on 0$\mathrm{^\circ}$ and 360$\mathrm{^\circ}$ longitude and reflection padding with 180$\mathrm{^\circ}$ rotation for the North and South Poles.

Spectral normalization is applied to all trainable FuXi layers. We found that this architectural change can stabilize the model training, especially in the multistep training stage. Detailed descriptions of spectral normalization are available in \citeA{schreck2024community}.

All changes above were implemented similarly in \citeA{schreck2024community} and referred to as ``CREDIT-FuXi''. The CREDIT-FuXi has reported competitive performance in \citeA{schreck2024community} and \citeA{sha2025improving}.

\noindent\textbf{S3. Full details of the hybrid sigma-pressure level conservation schemes}

This section provides technical explanations of the conservation schemes designed for hybrid sigma-pressure level AIWP models. The schemes constrain AIWP models on the conservation of global dry air mass, moisture budget, and total atmospheric energy. A similar set of conservation schemes proposed for constant pressure level models has been summarized and discussed in \citeA{sha2025improving}. Physical constants used in this section are summarized in Table~\ref{tab2}.

\begin{table}\label{tab2}
\begin{center}
\caption{Physical constants.}
\renewcommand{\arraystretch}{1.2}
\begin{tabularx}{\textwidth}
{>{\centering\arraybackslash}X c c c}
\specialrule{1.5pt}{0pt}{3pt}
Name & Abbreviation & Value & Units \\ 
\midrule
Radius of Earth & $R$ & 6371000 & $\mathrm{m}$ \\
Gravity of Earth & $g$ & 9.80665 & $\mathrm{m \cdot s^{-2}}$ \\
Density of Water & $\rho$ & 1000.0 & $\mathrm{kg \cdot m^{-3}}$ \\
Latent Heat of Vaporization\textsuperscript{a} & $L_v$ & $\mathrm{2.501 \times 10^6}$ & $\mathrm{J \cdot kg^{-1}}$ \\
Heat Capacity of Constant Pressure for Dry Air & $C_{pd}$ & 1004.64 & $\mathrm{J \cdot kg^{-1} \cdot K^{-1}}$ \\
Heat Capacity of Constant Pressure for Water Vapor & $C_{pv}$ & 1810.0 & $\mathrm{J \cdot kg^{-1} \cdot K^{-1}}$ \\
\specialrule{1.5pt}{3pt}{0pt}
\end{tabularx}
\end{center}
\vspace{1ex}
\textsuperscript{a} Value obtained on 273.15 K.
\end{table}

\noindent\textbf{S3.1 Preliminaries}

\noindent\textbf{S3.1.1 Global weighted sum}

For a given quantity $X(\phi, \lambda)$ that varies with latitude ($\phi$) and longitude ($\lambda$) as radians, its global weighted sum $\overline{X}$ in discrete form is defined as:

\begin{equation}\label{sec2_eq1}
\overline{X} = \text{SUM}\left(X\right) = \sum_{i_\phi=0}^{N_\phi} \sum_{i_\lambda=0}^{N_\lambda} {\left[X \cdot R^2 \cdot \Delta\left(\sin \phi \right) \cdot \Delta \lambda\right]}_{i_\phi,i_\lambda}
\end{equation}

\noindent
Where $R$ is the radius of the earth. $i_\phi=\left\{0, 1, \ldots, N_\phi\right\}$ and $i_\lambda=\left\{0, 1, \ldots, N_\lambda\right\}$ are indices of latitude and longitude grids, respectively. $\text{SUM}\left(X\right)$ is computed using the second-order difference for central grid cells and the forward difference for edge grid cells.

\noindent\textbf{S3.1.2 Hybrid sigma-pressure coordinate systems}

The sigma coordinates is a terrain-following coordinate system with its level defined based on surface pressure \cite{phillips1957coordinate}. An improved version of it, as used in the ECMWF-IFS and the main article, is the hybrid sigma-pressure coordinates.

The hybrid sigma-pressure coordinates follow the terrain near the surface and relax to constant pressure levels aloft. Its coordinate values are defined as follows:

\begin{equation}\label{sec2_eq2}
p_{i_l} = \alpha_{i_l} + \beta_{i_l}\cdot p_s
\end{equation}

\noindent
Where $p_s$ is surface pressure, $\alpha_{i_l}$ and $\beta_{i_l}$ are hybrid sigma-pressure level coefficients (hereafter, simplified as ``coefficients''). The hybrid sigma-pressure coordinate value ($p_{i_l}$) varies spatiotemporally with coefficients varying by levels and $p_s$ varying by spatial locations and time.

This study uses equation \ref{sec2_eq2} to describe the edge of each level (also known as the ``half-level''). Given $i_l=\left\{0, 1, \ldots, N_l\right\}$, when coefficients are indexed on $N_l$, their corresponding upper-air quantities, defined at grid centers are located between $N_l$ and $N_{l-1}$.

The hybrid sigma-pressure level AIWP configuration of this study has upper-air variables on 13 levels, which converts to coefficients on 14 half-levels, i.e., $N_l=14$. The values of these coefficients are summarized in Table~\ref{tab3}.

\begin{table}\label{tab3}
\begin{center}
\caption{The definition of hybrid sigma-pressure level coordinates.}
\renewcommand{\arraystretch}{1.2}
\begin{tabularx}{\textwidth}
{>{\centering\arraybackslash}X >{\centering\arraybackslash}X c >{\centering\arraybackslash}X}
\specialrule{1.5pt}{0pt}{3pt}
IFS level index\textsuperscript{a} & $\alpha$ [Pa] & $\beta$ & Half-level pressure [Pa]\textsuperscript{b} \\ 
\midrule
14 & 107.41574 & 0.000000 & 107.416 \\
15 & 131.4255 & 0.000000 & 131.425 \\
48 & 5119.895 & 0.000000 & 5119.895 \\
68 & 14432.14 & 0.007133 & 15154.891 \\
74 & 17467.613 & 0.026964 & 20199.741 \\
79 & 19343.512 & 0.059728 & 25395.451 \\
83 & 20219.664 & 0.099462 & 30297.651 \\
90 & 20087.086 & 0.203491 & 40705.812 \\
96 & 18006.926 & 0.332939 & 51741.970 \\
101 & 14898.453 & 0.466003 & 62116.207 \\
105 & 11901.34 & 0.576692 & 70334.655 \\
114 & 5564.383 & 0.790717 & 85683.783 \\
120 & 2659.1406 & 0.887408 & 92575.758 \\
133 & 62.78125 &  0.988500 & 100222.544 \\
\specialrule{1.5pt}{3pt}{0pt}
\end{tabularx}
\end{center}
\vspace{1ex}
\textsuperscript{a} This is the definition of 14 half levels. Variables are defined on the full levels.\\
\vspace{-2.5ex}
\textsuperscript{b} Computed using surface pressure of 101325 Pa.
\end{table}

\noindent\textbf{S3.1.3 Pressure level integral}

For a quantity $X(z)$ that varies with height $z$, its mass-weighted vertical integral can be converted to a pressure level integral using the hydrostatic equation:

\begin{equation}\label{sec2_eq3}
\int_{0}^{\infty}{\rho X}dz = \frac{1}{g}\int_{p_s}^{0}Xdp \approx \frac{1}{g}\int_{p_s}^{p_0}Xdp
\end{equation}

\noindent
Where $g$ is gravity, $p_0$ is the pressure of the model top.

Given hybrid sigma-pressure coordinate with half-level coefficients, the discrete form of equation \ref{sec2_eq3} is:

\begin{equation}\label{sec2_eq32}
\frac{1}{g}\int_{p_s}^{p_0}Xdp \approx \sum_{i_l=0}^{N_{l-1}}{\Delta p_{i_l} X_{i_l}},\quad \Delta p_{i_l} = p_{i_l} - p_{i_l-1}
\end{equation}

\noindent
Where $p_{i_l}$ and $p_{i_l-1}$ are half-level pressures computed using equation \ref{sec2_eq2}.

\noindent\textbf{S3.2 Global dry air mass conservation scheme}

As described in \citeA{sha2025improving}, the total amount of global dry air mass ($\overline{M_d}$) is conserved regardless of time. Given two forecast steps $\Delta t = t_1 - t_0$ with $t_0$ representing the analyzed initial condition and $t_1$ representing an arbitrary forecasted time, the conservation of $\overline{M_d}$ can be expressed as:

\begin{equation}\label{sec2_eq4}
\begin{aligned}
\overline{M_d} = \text{SUM}\left[\frac{1}{g}\int_{p_s}^{p_0}{\left(1-q\right)}dp\right]\\[6pt] 
\frac{\partial}{\partial t}\overline{M_d} = \overline{M_d\left(t_0\right)} - \overline{M_d\left(t_1\right)} = \epsilon_d
\end{aligned}
\end{equation}

\noindent
Where $q$ is atmospheric moisture, simplified using specific total water. $\epsilon_d$ is the residual term that violates the global dry air mass conservation.

In \citeA{sha2025improving}, specific total water was adjusted to force $\epsilon_d=0$. Here, a different approach is applied for hybrid sigma-pressure levels, with $p_s$ corrected to close the conservation budget. 

For this correction, the contribution of global dry air mass from coefficients $\alpha$ and $\beta$ are estimated as follows:

\begin{equation}\label{sec2_eq5}
\begin{aligned}
\overline{M_\alpha} = \text{SUM}\left[\frac{1}{g}\sum_{i_l=0}^{N_l-1}{\Delta\alpha_{i_l}\left(1-q\right)_{i_l}}\right],\quad \Delta\alpha_{i_l} = \alpha_{i_l} - \alpha_{i_l-1}\\[6pt] 
\overline{M_\beta} = \text{SUM}\left[\frac{p_s}{g}\sum_{i_l=0}^{N_l-1}{\Delta\beta_{i_l}\left(1-q\right)_{i_l}}\right],\quad \Delta\beta_{i_l} = \beta_{i_l} - \beta_{i_l-1}
\end{aligned}
\end{equation}

\noindent
Where $\overline{M_\alpha}$ and $\overline{M_\beta}$ are global dry air mass components spread to $\alpha$ and $\beta$, respectively. When computed on $t_1$, they are denoted as $\overline{M_\alpha\left(t_1\right)}$ and  $\overline{M_\beta\left(t_1\right)}$.

The correction of $p_s$ is defined as follows:

\begin{equation}\label{sec2_eq6}
p_s^*\left(t_1\right) = p_s\left(t_1\right)\frac{\overline{M_d\left(t_0\right)} - \overline{M_\alpha\left(t_1\right)}}{\overline{M_\beta\left(t_1\right)}}
\end{equation}

\noindent
Where $\overline{M_d\left(t_0\right)}$ is the total amount of global dry air mass calculated from the initial condition. $p_s^*\left(t_1\right)$ is the corrected $p_s$ on $t_1$. The same multiplicative correction ratio is applied to $p_s$ on all grid cells.

\noindent\textbf{S3.3 Global moisture budget conservation scheme}

The conservation of moisture budget in hybrid sigma-pressure level is consistent with \citeA{sha2025improving}. Here, the technical steps are summarized briefly. For more details, see \citeA{sha2025improving}.

The global sum of column-wise precipitable water ($M_v$) tendency is balanced by its sources and sinks. Namely, the global weighted sum of total precipitation ($\overline{P}$) and evaporation ($\overline{E}$):

\begin{equation}\label{sec2_eq7}
\begin{aligned}
\frac{\partial M_v}{\partial t} = \frac{1}{g}\frac{\partial}{\partial t}\int_{p_s}^{p_0}{q}dp=-E-P\\[6pt]
-\overline{\left(\frac{\partial M_v}{\partial t}\right)} - \overline{E} - \overline{P} = \epsilon_m
\end{aligned}
\end{equation}

\noindent
Where $\epsilon_m$ is the residual term that violates the global moisture budget conservation. $P$ and $E$ have units of $\mathrm{kg\cdot m^{-2} \cdot s^{-1}}$ with downward as positive.

$P$ is adjusted to ensure $\epsilon_m=0$ using a multiplicative ratio:

\begin{equation}\label{sec2_eq8}
P^*\left(t_1\right) = P\left(t_1\right)\frac{\overline{P^*\left(t_1\right)}}{\overline{P\left(t_1\right)}}, \quad\overline{P^*\left(t_1\right)} = -\overline{\left[\frac{M_v\left(t_1\right) - M_v\left(t_0\right)}{\Delta t}\right]} - \overline{E\left(t_1\right)}
\end{equation}

\noindent
Where $\overline{P^*\left(t_1\right)}$ is the corrected global sum of total precipitation required to close the moisture budget. The same multiplicative ratio is applied to all grid cells.

\noindent\textbf{S3.4 Global total atmospheric energy conservation scheme}

The conservation of global total atmospheric energy is consistent with \citeA{sha2025improving}. For a given air column, its vertically integrated total atmospheric energy ($A$) is defined as follows:

\begin{equation}\label{sec2_eq9}
A = \frac{1}{g}\int_{p_s}^{p_0}{\left(C_pT+L_v q+\Phi_s+k\right)}dp
\end{equation}

\noindent
The terms on the right side of the equation (\ref{sec2_eq9}) are thermal energy, latent heat energy, potential energy, and kinetic energy, respectively. $L_v$ is the latent heat of vaporization, and $\Phi_s$ is the geopotential at the surface. Kinetic energy ($k$) is defined as $k=0.5\left(\mathbf{v} \cdot \mathbf{v}\right)$. The specific heat capacity of air at constant pressure ($C_p$) is defined as $C_p=C_{pd}(1-q)+C_{pv}q$.

The formulation of $A$ in equation \ref{sec2_eq9} has some limitations due to the simplified use of moisture components. We have concluded that such limitations do not impact the use of conservation schemes for medium-range forecasts. Detailed discussion is available in \citeA{sha2025improving}. 

The global sum of column-wise total atmospheric energy tendency is balanced by its net energy sources and sinks on the top of the atmosphere ($R_T$) and the surface ($F_S$):

\begin{equation}\label{sec2_eq10}
\begin{aligned}
\overline{R_T} &- \overline{F_S} -\overline{\left(\frac{\partial A}{\partial t}\right)} = \epsilon_A \\[6pt]
R_T &= \mathrm{TOA}_{\mathrm{net}} + \mathrm{OLR} \\[6pt]
F_S &= R_{\mathrm{short}} + R_{\mathrm{long}} + H_s + H_l
\end{aligned}
\end{equation}

\noindent
Where $\epsilon_A$ is the residual term that violates the global total atmospheric energy conservation. $\mathrm{TOA}_{\mathrm{net}}$ is the top-of-atmosphere net solar radiation, $\mathrm{OLR}$ is outgoing longwave radiation. $R_{\mathrm{short}}$, $R_{\mathrm{long}}$, $H_s$, and $H_l$ are the surface net solar radiation, surface net longwave radiation, surface net sensible heat flux, and surface net latent heat flux, respectively. Frictional heating is ignored in $F_S$.

The air temperature ($T$) can be corrected to ensure thermal energy ($C_pT$) closes the energy budget, forcing $\epsilon_A=0$:

\begin{equation}\label{sec2_eq15}
\begin{aligned}
\overline{A^*\left(t_1\right)} = \overline{A\left(t_0\right)} + {\Delta t}\left(\overline{R_T} - \overline{F_S}\right), \quad \gamma = \frac{\overline{A^*\left(t_1\right)}}{\overline{A\left(t_1\right)}} \\
T^*\left(t_1\right) = \gamma T\left(t_1\right) + \frac{\gamma-1}{C_p}\left[L_v q\left(t_1\right)+\Phi_s+k\left(t_1\right)\right]
\end{aligned}
\end{equation}

\noindent
Where $\overline{A^*\left(t_1\right)}$ is the corrected global sum of total atmospheric energy, $\gamma$ is the multiplicative correction ratio. The same $\gamma$ is applied to $T$ at all grid cells and pressure levels.

\noindent\textbf{S4. Full details of model training}

This section summarizes the training routine and training objectives of the FuXi configurations. 

\noindent\textbf{S4.1 Training routines}

The FuXi configurations are trained in three stages: warm-up, single-step training, and multistep training. The warm-up stage has 10 epochs, with 100 batches per epoch and 32 samples per batch. The AdamW optimizer is used with weight decay of $3\times 10^{-6}$. The learning rate of this stage increases linearly from $3\times 10^{-7}$ to $1\times 10^{-3}$. The single-step training consists of 170 full epochs, with roughly 1800 batches per epoch. The same AdamW optimizer is used with an initial learning rate of $10^{-3}$, and the half cosine-annealing schedule. For multi-step training, the same AdamW optimizer is used, but with a fixed learning rate of $3\times 10^{-7}$. The multistep training consists of 12 full epochs. For each epoch, the number of iterative steps increases from 2 to 12, corresponding to forecast lead times of 12 to 72 hours. 

The above training stages are conducted on 32 NVIDIA A100 GPUs using Pytorch \cite{paszke2019pytorch} with fully sharded data parallel and activation checkpointing. The time cost of single-step training is roughly 50-60 minutes. The time cost of multi-step training varies by the number of steps. For 11-step (72-hour), it is roughly 14-16 hours per epoch. Hybrid sigma-pressure level and constant pressure level configurations have comparable training speeds, but using conservation schemes can cause a minor slowdown.

\noindent\textbf{S4.2 Training objective}

The training objective of all AIWP models is defined as follows:

\begin{equation}\label{equ_loss}
\mathcal{L}_{\mathrm{MSE}} = \frac{1}{N}\sum_{n=0}^N \frac{1}{l}\sum_{t=0}^l \frac{1}{G}\sum_{i=0}^G \frac{1}{V}\sum_{j=0}^V \left[a(i)s(j)w(j) \left(X_{i,j}^{n, t} - Y_{i,j}^{n,t}\right)\right]
\end{equation}

\noindent
Where $n=\left\{0, 1, \cdots N\right\}$ is the index of training batches. In this study, $N=32$. $l=\left\{0, 1, \cdots L\right\}$ is the index of forecast lead time. For single-step training, $l=L=0$. For multi-step training, $l$ ranges from 0 to 11. $i=\left\{0, 1, \cdots G\right\}$ is the index of grid cells. In this study, $G=181\times 360$. $j=\left\{0, 1, \cdots V\right\}$ is the index of variables. In this study $V=77$ for constant pressure level; $V=84$ for hybrid sigma-pressure levels. $a\left(i\right)$ is the latitude-based weighting. It is defined as $a\left(i\right)=\cos(\phi_i)$, $\phi_i$ is the latitude of each grid cell. This training objective is similar to \citeA{lam2023learning} and \citeA{sha2025improving}.

$s\left(j\right)$ is the per-variable-level inverse variance weights. It uses the tendency of the time difference of z-scored variables. Given a z-scored variable $X_j$, its tendency is calculated as:

\begin{equation}
\Delta X_j = \left\{ X_j(t=1) - X_j(t=0), \ldots, X_j(t+1) - X_j(t) \right\}
\end{equation}

\noindent
Where $t$ is the index of time. 

Per-variable-level inverse variance weights are the inverse of the standard deviation ($\sigma$) of $\Delta X_j$:

\begin{equation}
s\left(j\right) = \frac{1}{\sigma\left(\Delta X_j\right)}
\end{equation}

\noindent
Per-variable-level inverse variance weights are higher for variables that vary more strongly over space than over time, such as surface pressure and mean sea level pressure. These variables converge slowly during training. Implementing $s\left(j\right)$ can help resolve this problem. Per-variable-level inverse variance weights are applied to all output variables.

$w\left(j\right)$ is per-variable-level loss weight. It is related to the vertical level of a given variable. For constant pressure level configurations, $w\left(j\right)$ is defined as follows: 

\begin{enumerate}
    \item Upper-air variables below 300 hPa: $w=0.169$
    \item Upper-air variables on $\left\{1, 50, 150, 200, 250, 300\right\}$ hPa levels:
    \[
        w=\left\{1.69\times 10^{-4}, 0.00844, 0.0253, 0.0337, 0.0422, 0.105\right\}
    \]
    \item Prognostic single-level variables: $w=0.170$
    \item Diagnostic variables: $w=0.100$
\end{enumerate}

For hybrid sigma-pressure level configurations, $w\left(j\right)$ is defined as follows: 

\begin{enumerate}
    \item Upper-air variables from the 8th level to the 13th level: $w=0.179$
    \item Upper-air variables from the 1st level to the 7th level:
    \[
        w=\left\{1.82\times 10^{-4}, 0.00459, 0.0180, 0.0315, 0.0407, 0.050, 0.114\right\}
    \]
    
    \item Prognostic single-level variables: $w=0.2$
    \item Diagnostic variables: $w=0.1$
\end{enumerate}

\noindent
The constant pressure level per-variable-level weights were also used in \citeA{sha2025improving}. Figure~1a and b of the main article compared the relative height and weights of the two coordinate systems.

%
%

\section*{Open Research Section}
The ERA5 reanalysis data for this study can be accessed through the NSF NCAR Research Data Archive at \url{https://rda.ucar.edu/datasets/d633000/} and the Google Research, Analysis-Ready, Cloud Optimized (ARCO) ERA5 at \url{https://cloud.google.com/storage/docs/public-datasets/era5}. The IMERG Final Precipitation L3 daily product (GPM\_3IMERGDF) is obtained from the Goddard Earth Science Data and Information Science Center (GES-DISC), NASA at \url{https://disc.gsfc.nasa.gov/datasets/GPM_3IMERGDF_07/summary}. The IFS-HRES forecasts of this study are obtained from Weatherbench2; they are available at \url{https://weatherbench2.readthedocs.io/en/latest/data-guide.html}. The neural networks described here and the simulation code used to train and test the models are archived at \url{https://github.com/NCAR/miles-credit}. The verification and data visualization code of this study is archived at \url{https://github.com/yingkaisha/CREDIT-sigma-run}. 

\section*{Acknowledgments}
This material is based upon work supported by the National Science Foundation (NSF) National Center for Atmospheric Research (NCAR), which is a major facility sponsored by the U.S. National Science Foundation under Cooperative Agreement No. 1852977. This research has also been supported by NSF Grant No. RISE-2019758. We would like to acknowledge high-performance computing support from Derecho and Casper \cite{Cheyenne} provided by the Computational and Information Systems Laboratory, NCAR, and sponsored by the NSF.


\bibliographystyle{apacite}
\bibliography{reference}

%
%
%
%
%
\end{document}